\documentclass[twocolumn,showpacs,preprintnumbers]{revtex4}

\input{epsf}

\def\AJ{{\it Ap. J.} }

\def\ASAS{{\it Astron. and Astrophys.} }

\def\GRG{{\it Gen. Relativity and Gravitation} }

\def\MNRAS{{\it Mon. Not. R. Ast. Soc.} }

\def\PL{{\it Phys. Lett.} }
\def\PR{{\it Phys. Rev.} }

\def\al{\alpha}
\def\be{\beta}

\def\de{\delta}

\def\th{\theta}

\def\rh{\rho}

\def\Om{\Omega}

 \def\frac#1#2{{\textstyle{{#1}\over
{#2}}}} 
\def\lsim{\mathrel{\rlap{\lower4pt\hbox{\hskip1pt$\sim$}}
    \raise1pt\hbox{$<$}}} \def\gsim{\mathrel{\rlap{\lower4pt\hbox{\hskip1pt$\sim$}}
    \raise1pt\hbox{$>$}}}
\def\sqr#1#2{{\vcenter{\vbox{\hrule height.#2pt
         \hbox{\vrule width.#2pt height#1pt \kern#1pt
         \vrule width.#2pt}
         \hrule height.#2pt}}}} 

\def\beq{\begin{equation}}
\def\eeq{\end{equation}}
\def\beqa{\begin{eqnarray}}
\def\eeqa{\end{eqnarray}}

\begin{document}

\title{The Chaplygin dark star}

\vskip 0.2cm

\author{O. Bertolami, J. P\'aramos}

\vskip 0.2cm

\affiliation{Instituto Superior T\'ecnico, Departamento de
F\'{\i}sica, \\ Av. Rovisco Pais 1, 1049-001 Lisboa, Portugal}

\vskip 0.2cm

\affiliation{E-mail addresses: orfeu@cosmos.ist.utl.pt;
x\_jorge@fisica.ist.utl.pt}

\vskip 0.5cm

\date{\today}

\begin{abstract}
We study the general properties of a spherically symmetric body
described through the generalized Chaplygin equation of state. We
conclude that such object, dubbed generalized Chaplygin dark star,
should exist within the context of the generalized Chaplygin gas
model of unification of dark energy and dark matter, and derive
expressions for its size and expansion velocity. A criteria for
the survival of the perturbations in the GCG background that give
origin to the dark star are developed, and its main features are
analyzed.

\vskip 0.5cm

\end{abstract}

\pacs{98.80.-k, 97.10.-q \hspace{2cm} Preprint DF/IST-6.2005}

\maketitle

\section{Introduction}

The generalized Chaplygin gas (GCG) model has lately drawn some
attention, mainly because it allows for a unified description of
dark matter and dark energy, the first dominating at early times
and gradually transferring energy to the dark energy component
\cite{Kamenshchik, Bertolami1}. The GCG model is consistent with
various classes of cosmological tests, such as the Cosmic
Microwave Background Radiation \cite{Bertolami2}, supernovae
\cite{Bertolami3}, gravitational lensing \cite{Bertolami4} and
gamma-ray bursts \cite{Bertolami5}. As with other competing
candidates to explain the overwhelming energy density of the
present Universe, GCG is naturally constrained through
cosmological observables.

It is quite interesting that the GCG equation of state is that of
a polytropic gas \cite{Bhatia}, although one with a negative
polytropic index. This hints that one could look for astrophysical
implications of the model, and hence hope for yet another approach
to the problem of constraining the allowed space for its
parameters (see, e.g. Ref. \cite{Paramos}). In this work we argue
that a GCG dark star may arise from a density fluctuation in the
cosmological GCG background. In what follows we shall characterize
these objects, look at their evolution and account for their
initial probability of appearance within the GCG background.

\section{The generalized Chaplygin gas}

The GCG model is based on the equation of state

\beq P_{ch} = -{A \over \rho_{ch}^\al}~~, \label{state} \eeq

\noindent where $A$ and $\al$ are positive constants and $0 \leq
\al \leq 1$ (see however Ref. \cite{Bertolami7} for reasons to
consider $\al >1$); the negative pressure hints that the GCG is
related to a cosmological constant. The case $\al=1$ corresponds
to the Chaplygin gas \cite{Chaplygin}. In a
Friedmann-Robertson-Walker cosmology, the relativistic energy
conservation yields

\beq \rho_{ch} = \left[ A + {B \over a^{3(1+\al)} } \right]^{1
\over 1 + \al}~~, \label{rhoch} \eeq

\noindent where $a$ is the scale factor of the Universe and $B$ a
positive integration constant. This result shows a striking
property of the GCG, namely that at early times it behaves as
non-relativistic dark matter ($\rho_{ch} \propto a^{-3}$), while
at late times it acts as a cosmological constant ($\rho_{ch}
\simeq const.$). One can algebraically decompose the GCG into a
dark matter and a dark energy component, which evolve such that
the transferring of energy occurs from the former to latter
\cite{Bertolami6}. This can be used to show that, while the dark
energy component is spatially homogeneous, the dark matter
component allows for structure formation at early times, when it
dominates \cite{Bertolami1,Bertolami6,Bilic}.

For convenience, one defines the parameter $ A_s \equiv A /
\rho_{ch0}^{1+\al}$, where $\Om_{de0}$ is the dark energy density,
$\rho_{cr0}$ the critical density and $\rho_{ch0}$ the GCG energy
density, all at the present. Assuming, as observations suggest,
the condition $\Om_{dm0} + \Om_{de0} = 1$, where $\Om_{dm0}$ is
the dark matter density (dropping the small baryon contribution),
and taking $a_0=1$, yields the constraint $ B = \Om_{dm0}
\rho_{cr0} \rho_{ch0}^\al$.

\section{Polytropic stars}

In order to deal with stellar structure in general relativity, one
considers that the spherical body behaves as a perfect fluid,
characterized by the energy-momentum tensor

\beq T^{\mu \nu} = (\rho + P)u^\mu u^\nu + P g^{\mu \nu} ~~,\eeq

\noindent where $g^{\mu\nu}$ is a spherically symmetric Birkhoff
metric. The Bianchi identity implies that $\nabla_\mu T^{\mu \nu}
= 0$, from which follows the relativistic
Tolman-Oppenheimer-Volkov equation \cite{Bhatia},

\beq {d P \over d r } = -{G( P + \rho) \over r^2} \left[m+4 \pi
r^3 P \right]\left[1 - {2G m\over r} \right]^{-1}~~, \label{TOV}
\eeq

\noindent where $ m(r) = 4 \pi \int_0^r \rho(r') r'^2 dr'$. This
equation collapses to the classical Newtonian hydrostatic
equilibrium equation

\beq {d P \over dr } = -{4 G m(r) \rho \over r^2} ~~,
\label{hydro} \eeq

\noindent if the following conditions are satisfied:

\beqa && G m(r) /r \ll 1~~,  \label{newta} \\ && 4 \pi r^3 P(r)
\ll m(r) ~~, \label{newtb} \\ && P(r) \ll \rho(r)~~. \label{newtc}
\eeqa

The polytropic gas model for stellar like structure assumes an
equation of state of the form $P = K \rho^{n+1/n}$, where $n$ is
the polytropic index, which defines intermediate cases between
isothermic and adiabatic thermodynamical processes, and $K$ is the
polytropic constant, defined as

\beq K = N_n G M^{(n-1)/n} R^{(3-n)/n} ~~ \label{k} \eeq

\noindent with

\beq N_n = \left[{n+1 \over (4\pi)^{1/n}} \xi^{(3-n)/n}
\left(-\xi^2 {d \th \over d \xi}
\right)^{(n-1)/n)}\right]^{-1}_{\xi_1}~~,\eeq

\noindent where $R$ is the star's radius, $M$ its mass and
$\xi_1$, defined by $\th(\xi_1) \equiv 0$, corresponds to the
surface of the star (cf. below). Actually, this definition states
that all quantities tend to zero as one approaches the surface.

This assumption leads to several scaling laws for the relevant
thermodynamical quantities,

\beqa \rho & = & \rho_c \th^n(\xi)~~,~~~~ \label{defrho} \\
T & = & T_c \th(\xi)~~,~~~~ \\ P & = & P_c \th(\xi)^{n+1}~~,~~~~
\label{defP} \eeqa

\noindent where $\rho_c$, $T_c$ and $P_c$ are the density,
temperature and pressure at the center of the star \cite{Bhatia}.
Notice that the scaling law for temperature requires the
assumption that the gas behaves as an ideal one. This is not the
case of the GCG.

The function $\th$ is a dimensionless function of the
dimensionless variable $\xi$, related to the physical distance to
the star's center by $r = \be \xi$, where

\beq \be = \left[{(n+1)K \over 4 \pi
G}\rho_c^{(1-n)/n}\right]^{1/2} ~~. \label{beta} \eeq

\noindent The function $\th(\xi)$ obeys a differential equation
arising from the equilibrium condition of Eq.(\ref{hydro}), the
Lane-Emden equation:

\beq {1 \over \xi^2} {\partial \over \partial \xi} \left(\xi^2
{\partial \th \over \partial \xi} \right) = -\th^n~~. \label{lec}
\eeq

\noindent Notice that the physical radius and mass of the
spherical body appear only in the polytropic constant, and hence
the behavior of the scaling function $\th(\xi)$ is unaffected by
these. Therefore, the stability of a star is independent of its
size or mass, and different types of stars correspond to different
polytropic indices $n$. This scale-independence manifests in the
symmetry of the Lane-Emden equation (\ref{lec}) under homology
transformations. The first solar model ever considered, developed
by Eddington in 1926, was that of an $n=3$ polytropic star.
Although somewhat incomplete, this simplified model gives rise to
relevant constraints on the physical quantities.

In what follows, we shall use the Lane-Emden equation to derive
the properties of a generalized Chaplygin dark star, given that
conditions (\ref{newta})-(\ref{newtc}) are shown to be fulfilled.
This enables the use of the Newtonian approximation (\ref{hydro}),
which asides its simplicity allows for a prompt interpretation of
the GCG as a polytropic gas subject to the Lane-Emden equation of
motion. The generality of this procedure can be used in various
cases of physical interest, as for instance, when studying the
effect of scalar fields on the stellar structure \cite{Paramos}.

\section{The generalized Chaplygin dark star}

As already discussed, the GCG model is cosmological in nature, and
most bounds on the parameters $\al$ and $A_s$ are derived from
cosmological tests. However, a quick look at the GCG equation of
state (\ref{state}) indicates that the it corresponds to a
polytrope with a negative polytropic constant and a negative
pressure. At first glance, this seems to indicate that no valid
analysis of a GCG at an astrophysical context can proceed, since a
spherical body constituted by such exotic gas would experience an
outward pressure that would prevent it from being stable. However,
the following argument shows that such an objection is
circumvented by the presence of the cosmological GCG background.

The first logical step for the construction of a symmetric body
with the GCG equation of state should be, as for all polytropes,
the solution of the related Lane-Emden equation. Firstly one notes
that this stems from the hydrostatic equation (\ref{hydro}). As
already seen, this is directly derived from the general relativity
equations, assuming the energy-momentum tensor of a perfect fluid:
no assumption whatsoever is made concerning the pressure or
density, nor the equation of state relating these quantities.
Hence, one can use this equation and, through the usual
derivation, the related Lane-Emden equation; as stated before the
only concern is if one can neglect the higher-order relativistic
terms present in Eq.(\ref{TOV}), thus working in the Newtonian
limit. This will be explicitly shown in the case under
investigation.

The polytropic equation of state $P = \rho^{n+1/n}$ shows that the
GCG can be assigned a negative polytropic index $n = -1/(1 +
\al)$. Next, a comparison with Eq. (\ref{state}) yields $K = - A$,
since the pressure is negative. This requires some caution:
indeed, the direct application of the coordinate transformation
between the physical radial coordinate $r$ and the coordinate
$\xi$ given by by $r = \be \xi$, with $\be$ defined in Eq.
(\ref{beta}) yields

\beq \be \equiv D \left[ ( 1 + n) K \right]^{1/2} = D \left( {\al
\over 1 + \al} K \right)^{1/2} ~~, \eeq

\noindent where $D=[\rho_c^{(1-n)/n} / 4 \pi G]^{1/2}$; since $\al
> 0$ and $K < 0$, the above quantity is imaginary. To avoid this, we
define the coordinate $\xi$ through the same equation $r = \be
\xi$, but with $K$ replaced by $|K| = A > 0$ in Eq. (\ref{beta}),
obtaining

\beq \be = \left[{A \over 4 \pi G } {\al \over 1 + \al}
\right]^{1/2} \rho_c^{-(1+\al/2)}~~. \label{betag} \eeq

The negative sign of $K$ will, of course, manifest itself in the
terms of the Lane-Emden equation; explicitly, one gets

\beq {1 \over \xi^2} {\partial \over \partial \xi} \left(\xi^2
{\partial \th \over \partial \xi} \right) = \th^{n}~~. \label{leg}
\eeq

As in the usual Lane-Emden equation, one has as boundary
conditions $\th(0)=1$ and $\th'(0)=0$. This gives rise to a
positive derivative for $\xi>0$, indicating that $\th$ is a
smoothly increasing function. This is related not to the negative
polytropic index, but to the negative pressure of the GCG; as a
consequence, Eqs. (\ref{defrho}) and (\ref{defP}) indicate that
the pressure inside the dark star increases (in absolute value),
while the density decreases. This is key to our study, since it
shows that a GCG spherical body accretes, as expected for a star.

In the usual Lane-Emden equation, the criteria concerning the size
of a star is given by $\th(\xi_1) \equiv 0$, corresponding to a
surface of zero density, pressure and temperature. In a GCG dark
star, the question is more convoluted: a vanishing density yields
infinite pressure (and conversely), which are rather unphysical
choices for the boundary of any astrophysical object. Furthermore,
since the function $\th$ is increasing, the density $ \rho \propto
\th^{n}$ vanishes at an infinite distance, while the pressure $P
\propto - \th^{1+n}$ does not vanish at all.

\begin{figure}

\epsfysize=6.8cm \epsffile{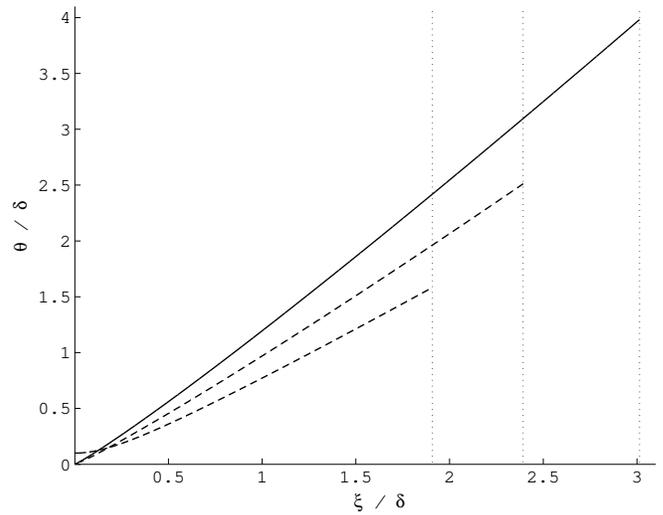} \caption{The function
$\th(\xi) / \de $ for a relative density $\de = 10$, $100$, $1000$
(dashed, dot-dashed and full lines respectively) as a function of
$\xi / \de$, assuming $\al=0.2$ and $A_s=0.7$.} \label{fig}

\end{figure}

As a solution for this issue, one recalls that the GCG object is
embedded in a cosmological background. Hence, a GCG dark star
should not be taken as an isolated body, but rather as a spike on
the overall cosmological background of density $\rho_{ch}$ and
negative pressure $P_{ch}$. Therefore, its boundary should be
signalled by the matching of the inner and outer pressures and
densities, as indicated in Fig. \ref{fig}. Both conditions are, of
course, equivalent, given the GCG equation of state (\ref{state}).
From Eqs. (\ref{defrho}) and (\ref{defP}), this equates to

\beq \th(\xi_1) \equiv \de^{-1/n} =  \de^{1+\al} ~~, \label{x1}
\eeq

\noindent where one defines the ratio between central and the
background density, $\de \equiv \rho_c / \rho_{ch}$. Hence, one
gets a correspondence between the central density of the dark
star, that is, the height of the energy density fluctuation, and
its radius. This argument shows that the Chaplygin dark star
cannot be taken merely as an isolated body having a common
equation of state with the GCG, but must be viewed instead as a
perturbation to the flat GCG background. This is advantageous,
since one can use the constraints available on the GCG model to
ascertain its properties. The only constraint affecting this
quantity is $\de \gg 1$, since we assume that the density
perturbation must be large enough so we can view it as a physical
object, not merely a small fluctuation on the GCG background.

Notice that, since the scaling function $\th(\xi)$ is completely
specified by the two boundary conditions $\th(0)=1$ and
$\th'(0)=0$, neither $\rho_c$ nor $\rho_{ch}$ affect each other:
$\rh_c$ is ``scaled out'' of the problem through the definition
(\ref{defrho}) and(\ref{defP}) and $\rho_{ch}$ merely sets the
criteria for the surface of the star, through Eq. (\ref{x1}).
Hence, although $\rho_{ch}$ varies with time, there is no
contradiction in assuming a constant central density $\rho_c$;
this simplifies our study, since Eq. (\ref{betag}) then yields a
constant coordinate scaling coefficient $\be$. Furthermore, given
that the cosmological background density $\rho_{ch}$ is
decreasing, a constant central density $\rho_c$ indicates that the
density ratio $\de$ increases and, therefore, $\xi_1$ expands
towards a final radius $r_{1 \infty} = \be \xi_{1 \infty}$,
acquiring mass in the process. One can argue that, due to energy
conservation, the background density should decrease in order to
compensate this, but this is a minor effect that can be neglected.
Also, since the Chaplygin dark star is not an isolated object, it
is reasonable to assume that neither its mass nor radius should be
held constant, but instead must vary as it dilutes itself on the
overall cosmological background; instead, the central density
$\rho_c$ arises as the natural candidate for distinguishing
between these objects.

The absolute magnitude of the perturbation results from the
dynamics ruling the generation of a perturbation, via a
probability law arising from the fundamental physics underneath
the GCG model; this, of course, should naturally disfavor very
large perturbations. Furthermore, since any relative perturbation
to the homogeneous energy density profile is local, its occurrence
should not depend on cosmological quantities and the probability
distribution should depend only on the relative perturbation
$\de$, and not explicitly on the scale factor $a$. A putative
candidate could be a normal probability distribution given by

\beq f(\de) = f_0 \exp\left[-g \de^2 \right]~~, \label{prob} \eeq

\noindent where $f_0$ is a normalization factor, $g$ a parameter
dependent on $A_s$ and $\al$. Of course, more complicated
expressions for $f(\de)$ are possible, depending on the inner
workings of the fundamental physics behind the GCG model.

The expansion velocity of a dark star can be shown to be given by

\beqa v_e & \equiv & \dot{r}_1 = {3 (1+\al) \be B \over
\th'(\xi_1) } \left[{\rho_c \over \rho_{ch}^2} \right]^{1+\al} {H
\over a^{3(1 + \al})}~~, \label{vexp} \eeqa

\noindent where the prime denotes differentiation with respect to
$\xi$ and $H=\dot{a}/a$ is the rate of expansion of the
cosmological background. The Friedmann equation allows one to
write the latter as

\beq H^2 = H_0^2 \left[ \Om_{de0} + \Om_{dm0} a^{-3(1+\al)}
\right]^{1 \over 1 + \al}~~. \label{exprate} \eeq

\noindent One can see that, as $\rho_{ch}$ approaches a constant
value at late times, the expansion tends to zero. Also, one can
derive the dependence of $v_e$ on $\de$ by noticing that the
scaling coefficient $\be$ runs with $\de^{-(1+\al/2)}$ and the
term in brackets in Eq. (\ref{vexp}) scales with $\de$, while
$\th'(\xi_1)$ is found numerically to always be of order unity.
Hence, one concludes that the expansion velocity depends weakly on
$\de$, $v_e \propto \de^{-\al/2} \sim \de^{-0.1}$, given the
chosen value of $\al=0.2$ \cite{Bertolami6,Bertolami8}. By the
same token, since numerically one finds that $\xi_1 \propto \de$,
it can be concluded that $r_1 = \be \xi_1 \propto \de^{-\al/2}
\sim \de^{-0.1}$.

Given that the GCG tends to a smooth distribution over space, most
density perturbations tend to be flattened within a timescale
related to their initial size and the characteristic speed of
sound $v_s = (
\partial P /
\partial \rho )^{1/2}$. Inside the dark star, the equation of
state (\ref{state}) and the definitions (\ref{defrho}) and
(\ref{defP}) yield

\beq v_{s, in} = \sqrt{\al A \th(\xi) \over \rho_c^{1 + \al}}
\equiv \sqrt{\al A_s {\th(\xi) \over \th(\xi_1)}}
\left({\rho_{ch0} \over \rho_{ch}}\right)^{(1+\al)/2} ~~, \eeq

\noindent which at the surface amounts to

\beq v_{s, ch} = \sqrt{\al A \over \rho_{ch}^{1 + \al}} \equiv
\sqrt{\al A_s} \left({\rho_{ch0} \over
\rho_{ch}}\right)^{(1+\al)/2}~~. \label{vsurf} \eeq

\noindent One sees that the maximum sound velocity occurs at the
surface of the star; since $\al \lesssim 0.6$ and $ 0.6 \leq A_s
\leq 0.8$ (see first references in \cite{Bertolami2}), this should
be smaller than the present value of $v_{s, max} = 0.693$ (in
units of $c$). A plausible criteria for the survival of an initial
perturbation is given by $ v_{s, ch}< v_{e, 0}$, where $v_{e, 0}$
is the initial expansion velocity. Equating Eqs. (\ref{vexp}) and
(\ref{vsurf}) yields, after a little algebra

\beq \th'(\xi_1) \de^{-\al /2} < {3 \over 2} \sqrt{1 + \al \over
\pi G \rho_{ch} } \left[ H_0 \over H \right]^{1+ 2\al} {\Om_{dm0}
H_0 \over a^{3(1+\al)} } ~~. \label{cond} \eeq

\noindent Numerically, one finds that the left-hand side of Eq.
(\ref{cond}) is approximately constant. At early times, the GCG
behaves as cold dark matter, with $\rho_{ch} \propto a^{-3}$ and
$H \propto a^{-3/2}$, and the right hand side of condition
(\ref{cond}) is constant. At late times, when the GCG acts as a
cosmological constant, both $\rho_{ch} $ and the expansion rate
$H$ are constant, and the {\textit rhs} then scales as
$a^{-3(1+\al)}$, and thus decreases with cosmic time. Thus, most
dark stars are created at early times; this is consistent with the
usual interpretation of the GCG as a model where dark matter
dominates at early times, allowing for structure formation, while
at late times the dark energy component takes over.

Eq.(\ref{vsurf}) and the GCG equation of state (\ref{state})
allows one to rewrite condition (\ref{newtc}) as

\beqa & & \left|{P(r) \over \rho(r)}\right| = {A \over
\rho(r)^{1+\al}} = A_s \left(\rho_{ch0} \over \rho(r)
\right)^{1+\al} \\ \nonumber && < A_s \left(\rho_{ch0} \over
\rho_{ch} \right)^{1+\al} < 1 ~~, \eeqa

\noindent where we have used $\rho(r) > \rho_{ch} >
\rho_{ch\infty}$ for any redshift $z$, with $\rho_{ch\infty} =
A^{1/1+\al}$ the limit for the background cosmological density
when $a \rightarrow \infty$. Relativistic corrections could be
important if the pressure and density are of the same order of
magnitude. However, one can use the bound

\beq \left|{P(r) \over \rho(r)}\right| < A_s \left(\rho_{ch0}
\over \rho_{ch} \right)^{1+\al} ~~, \eeq

\noindent to ascertain that, for redshifts typical for structure
formation, $z=z_c=15$ and a set of GCG parameters $\al=0.2$,
$A_s=0.7$, one gets $|P(r) / \rho(r)| < 10^{-4}$; higher redshifts
provide an even lower upper bound. For a much recent $z = 1$, the
same set of parameters yields $|P(r) / \rho(r)| < 0.188$, which
still validates the Newtonian approximation, although to a lesser
extent. This is not troublesome, since most dark stars are assumed
to nucleate at an early age. Also, since the above condition only
provides an upper limit for $P(r) / \rho(r)$, a more complete
calculation can still validate the Newtonian approximation,
depending on the value of $\de$.

Assuming that all dark stars have expanded up to their final size
(which follows from the stabilization of the GCG as a dark
energy), one can write the mass contribution of those created when
the Universe had a size $a(t)$:

\beqa && {M(a) \over 4 \pi} = \int_0^{\rho_n} \int_0
^{r_1(\rho_c)} \rho(r,\rho_c) f(\rho_c) r^2 dr ~d\rho_c
\\ \nonumber && = \be^3 \int_0^{\rho_n} \int_0
^{\xi_1(\rho_c)} \rho_c \th^n(\xi,\rho_c) f(\rho_c) \xi^2 d \xi
~d\rho_c ~~, \eeqa

\noindent where $r_1 = \be \xi_1$ and the dependence on the
integration variables is made explicit. Integrating over time one
gets the mass contribution of all generations of dark stars, $
M_{DS} = \int_0 ^{a_0} M'(a) da $, where $M'(a) \equiv dM/da$.

A comparison between known observational bounds and numerical
integration of the above results can then be used to constraint
the GCG parameters $A_s$ and $\al$, namely through supernovae
data, gravitational lensing results and other dark matter
searches. This will be considered elsewhere.

\section{Numerical Results}

In order to substantiate our arguments, in this section we present
some numerical examples; we shall study the proposed scenario for
the ``typical'' values of the GCG model; one takes $\al=0.2$ and
$A_s=0.7$. A future study based on this results could embrace a
wider range of parameters and probe the creation of dark stars at
early stages of the Universe, providing further refinement to the
already known bounds (see, e.g. Ref. \cite{Bertolami8} for a
summary of the existing constraints).

A numerical integration of the modified Lane-Emden equation
(\ref{leg}) produces the results plotted in Fig. \ref{fig}. In
Table I we draw different scenarios, in order to ascertain the
dimensions of dark stars nucleated at different ages of the
Universe. One can see that at a redshift of $z=z_c=15$, presumably
typical for structure formation, even a small perturbation $\de=5$
produces an overwhelmingly large object, with about $3000$ times
the mass and $20$ times the diameter of the Milky Way. Since these
dimensions scale with $\be$, which decreases with $\rho_c$, one
probes higher redshifts in order to obtain smaller dark stars.
Therefore, at $z=50$ and $\de=10$, one obtains an object with
approximately the size and double the mass of our galaxy. A larger
perturbation $\de = 100$ yields approximately the same size, but a
ten-fold increase in mass.

Going further back in time, a darkstar born at $z=100$ with
$\de=10$ ($100$) has about one-hundredth (one-tenth) the mass of
the Milky Way and one-tenth its diameter. Finally, $\de=100$ and
an extremely high redshift of $z=500$, deep within the so-called
``dark ages'', yield a dark star with $1.6 \times 10^6$ solar
masses and a radius of $7.8~pc$, dimensions similar to those
ascribed to super-massive black holes in active galactic nuclei.

The above discussion is by no means definitive, and only serves to
illustrate the concept developed in this study. Nevertheless, one
might be surprised by the unphysically large size of a dark star
hypothetically nucleated at the redshift typical for structure
formation, $z_c=15$. However, notice that this describes the
condensation of baryonic matter interacting gravitationally with
dark matter. The dark star scenario poses quite a different
mechanism, where the GCG (in an era where its dark matter
component dominates) is the sole constituent of the spherical
body. Hence, it is reasonable to assume that bodies of
astrophysical dimensions can arise much earlier in the history of
the Universe. A precise description would of course imply the
nucleation probability distribution $f(\de)$, since this is the
fundamental quantity ruling the onset of perturbations nucleation
on the GCG background.

To ascertain the validity of the Newtonian limit from which the
Lane-Emden equation is derived, conditions (\ref{newta}) and
(\ref{newtb}) can now be checked with a simple calculation. An
inspection of the Table I shows that $\xi_1$ is of the order of
$\de$. Also, Fig. \ref{fig} shows that while $\th'(\xi_1)$ slowly
increases and is of order unity, the scaling function $\th(\xi)$
grows regularly; hence, one can use the power-law approximation
$\th(\xi) \sim a \xi^b $, where $a$ and $b$ are coefficients of
order unity, since it does not introduce large deviations from a
full numerical calculation. Using Eq. (\ref{defrho}), this yields

\beq m(r) \sim {4 \pi \over 3 + n b } \rho(r) r^3 ~~, \eeq

\noindent and condition (\ref{newta}) becomes

\beq {4 \pi \over 3 + n b } G \rho(r) r^2 \ll 1~~.
\label{newtaapprox} \eeq

\noindent Since $\rho(r) \propto r^{nb}$, the left-hand side of
Eq. (\ref{newtaapprox}) scales with $r^{2+n b}$ and, since $|n b|
\sim 1$, is an increasing function. Therefore, it is majored by
its value at $r=r_1$, amounting to $ \sim 4 \pi G \rho_{ch} r_1^2
$. Using Table I one finds that it attains a maximum value of
$5.12 \times 10^{-5}$ for $\de=5$, $z=z_c=15$, thus concluding
that condition (\ref{newta}) is verified. Notice that, for a fixed
redshift, this ratio is approximately independent of $\de$. This
is due to the very weak scaling of the physical radius with the
relative perturbation, $r_1 \propto \de^{-\al/2} \sim \de^{-0.1}$,
for the chosen value $\al=0.2$.

In a similar fashion, the power-law approximation allows one to
write condition (\ref{newtb}) as

\beq P(r) \ll \rho(r) {1 \over 3 + n b}~~, \eeq

\noindent which, since $|n b | \sim 1 $, is equivalent to
condition (\ref{newtc}) and hence also satisfied at early ages.
Thus, the Newtonian approximation implicit in the Lane-Emden
equation is valid for the cases studied and, given the smallness
of the values encountered in its evaluation, it is also applicable
in a broader range of the nucleation redshift $z$ and relative
densities $\de$.

Given the indicated values for the initial expansion velocity
$v_{e0}$ and the surface sound velocities $v_s$ (which depends
only on the redshift) the inequality (\ref{cond}) is valid for the
chosen values, and thus the corresponding dark stars do not
collapse at birth.

\begin{widetext}

\begin{table}
\begin{ruledtabular}

\begin{tabular}{|c|c|c|c|c|c|c|c|c|c|c|c|c|c|}

& $ \de = 5~,~ z=15 $ & & $\de = 10~,~z=50$ & & $\de = 100~,~z=50$ & & $\de = 10~,~z=100$ & & $\de = 100~,~z=100$ & & $\de = 100~,~z=500$ \\

\hline

$ \xi_1 $ & $ 9.04 $ & & $ 19.1$ & & $ 239 $ & & $ 19.1 $ & & $ 239 $ & & $ 239 $ \\
$ \th'(\xi_1) $ & $ 0.850 $ & & $ 0.916 $ & & $ 1.14 $ & & $ 0.916 $ & & $ 1.14 $ & & $ 1.14 $ \\
$ \rho_c~(Kg.m^{-3}) $ & $ 6.11\times 10^{-23 }$ & & $ 3.96\times 10^{-21 }$ & & $ 3.96\times 10^{-20 }$ & & $ 3.07\times 10^{-20 }$ & & $ 3.07\times 10^{-19 }$ & & $ 3.75\times 10^{-17 }$ \\
$ \be~ (pc) $ & $ 7.59 \times 10^4 $ & & $ 772 $ & & $ 61.4 $ & & $ 81.0 $ & & $ 6.44 $ & & $ 3.26 \times 10^{-2 }$ \\
$ r_1 ~(pc)$ & $ 6.86 \times 10^5 $ & & $ 1.48\times 10^4 $ & & $ 1.47\times 10^4 $ & & $ 1.55\times 10^3 $ & & $ 1.54\times 10^3 $ & & $ 7.81 $ \\
$ M / M_\odot$ & $ 1.73\times 10^{ 15 }$ & & $ 1.13\times 10^{ 12 }$ & & $ 1.11\times 10^{ 13 }$ & & $ 1.02\times 10^{ 10 }$ & & $ 9.97\times 10^{ 10 }$ & & $ 1.58\times 10^6 $ \\
$ 4 \pi G \rho_{ch} r_1^2 /c^2 $ & $ 5.12\times 10^{-5 }$ & & $ 7.66\times 10^{-7 }$ & & $ 7.58\times 10^{-7 }$ & & $ 6.55\times 10^{-8 }$ & & $ 6.48\times 10^{-8 }$ & & $ 2.03\times 10^{-10 }$ \\
$ v_e /c $ & $ 1.89\times 10^{-2 }$ & & $ 2.33\times 10^{-3 }$ & & $ 2.35\times 10^{-3 }$ & & $ 6.81\times 10^{-4 }$ & & $ 6.86\times 10^{-4 }$ & & $ 3.84\times 10^{-5 }$ \\
$ v_s/c$ & $ 5.09\times 10^{-3 }$ & & $ 6.32\times 10^{-4 }$ & & $ 6.32\times 10^{-4 }$ & & $ 1.85\times 10^{-4 }$ & & $ 1.85\times 10^{-4 }$ & & $ 1.03\times 10^{-5 }$ \\
$ \th'(\xi_1) \de^{-\al/2} $ & $ 0.724 $ & & $ 0.728 $ & & $ 0.722
$ & & $ 0.728 $ & & $ 0.722 $ & & $ 0.722 $

\label{table}

\end{tabular}

\caption{Numerical results for the quantities $\xi_1$,
$\th'(\xi_1)$, $\rho_c$, $\be$, $r_1$, $M$, $ 4 \pi G \rho_{ch}
r_1^2 /c^2 $, $v_{e0}$, $v_s$ and $\th'(\xi_1) \de^{-\al/2}$, for
$\de=10,~100$ and redshifts $z=15,~50,~100,~500$.}

\end{ruledtabular}
\end{table}

\end{widetext}

\section{Conclusions}

In this study we have analyzed the properties of spherical bodies
with a polytropic equation of state of negative index, in the
context of the GCG dark energy/dark matter unification model. We
have considered the associated Lane-Emden equation and looked at
the qualitative behavior of its solution; amongst the results we
find the conditions for fluctuations to be attenuated or to
develop as dark stars. Our criteria is based on the condition that
the sound velocity does not exceed the expansion velocity of the
dark star when it nucleates $v_{s, ch} < v_{e, 0}$. This enables
the computation of the mass contribution of the dark stars at
present times, which can then be used to constraint the GCG
parameters $A_s$ and $\al$, providing another testing ground for
this fascinating model.

\vskip 0.2cm

{\bf Note added:} While finalizing this work we became aware of
the study of stable dark energy objects \cite{Lobo} and halos of
$k$-essence \cite{Lim}. Even though the motivation of both works
are somewhat similar, our approaches are quite different.


\begin{acknowledgments}

\noindent JP is sponsored by the Funda\c{c}\~{a}o para a
Ci\^{e}ncia e Tecnologia under the grant BD~6207/2001.

\end{acknowledgments}

\end{document}